\documentstyle[twoside,fleqn,espcrc2,epsfig]{article}

\newcommand{\AmS}{{\protect\the\textfont2
  A\kern-.1667em\lower.5ex\hbox{M}\kern-.125emS}}

\def\eq#1{{eq. (\ref{#1})}}

\def\gsim{\;
\raise0.3ex\hbox{$>$\kern-0.75em\raise-1.1ex\hbox{$\sim$}}\;}
\def\lsim{\;
\raise0.3ex\hbox{$<$\kern-0.75em\raise-1.1ex\hbox{$\sim$}}\;}

\newcommand\beq{\begin{equation}}
\newcommand\eeq{\end{equation}}

  \def\gsim{\;
  \raise0.3ex\hbox{$>$\kern-0.75em\raise-1.1ex\hbox{$\sim$}}\; }
\def\lsim{\;
  \raise0.3ex\hbox{$<$\kern-0.75em\raise-1.1ex\hbox{$\sim$}}\; }

\newlength{\digitwidth} 
\catcode`?=\active \def?{\kern\digitwidth}
\title{Solar neutrino problem accounting
for self-consistent magnetohydrodynamics solution
for solar magnetic
fields\thanks{The talk given by T.I.Rashba at the EuroConference on Frontiers in Particle Astrophysics and Cosmology, San Feliu de Guixols, Spain, 30 September - 5 October, 2000. It based on the work of the authors hep-ph/0005259, which to be published in Nuclear Physics B.}}

\author{%
\underline{T.~I.~Rashba}%
\address{%
The Institute of the Terrestrial Magnetism,
the Ionosphere and Radio Wave Propagation of the Russian
Academy of Sciences, IZMIRAN, Troitsk, Moscow region, 142190, Russia}%
\thanks{E-mail: rashba@izmiran.rssi.ru},
O.~G.~Miranda%
\address{Departamento de F\'{\i}sica,
CINVESTAV-IPN, A. P. 14-740, M\'exico 07000, D. F., M\'exico.}%
\thanks{E-mail: omr@fis.cinvestav.mx},
C.~Pe\~na-Garay%
\address{%
Instituto de F\'{\i}sica Corpuscular -
C.S.I.C./Universitat de Val\`encia, Edificio Institutos de Paterna,
Apartado de Correos 2085, 46071 Val\`encia, Spain\\
{\tt http://neutrinos.uv.es}}%
\thanks{E-mail: penya@flamenco.ific.uv.es},
V.~B.~Semikoz$^{a}$%
\thanks{E-mail: semikoz@flamenco.ific.uv.es}
and
J.~W.~F.~Valle$^{c}$%
\thanks{E-mail: valle@flamenco.ific.uv.es},
}

\begin{document}

\begin{abstract}

  The analysis of the resonant spin--flavour (RSF) solutions to the
  solar neutrino problem in the framework of simplest analytic solutions
  to the solar magneto-hydrodynamics (MHD) equations is presented.
  We performed the global fit of the recent solar neutrino data,
  including event rates as well as day and night
  recoil electron spectra induced by solar neutrino interactions
  in SuperKamiokande.
  We compare quantitatively our simplest MHD-RSF fit with vacuum
  oscillation (VAC) and MSW--type (SMA, LMA and LOW) solutions to the
  solar neutrino problem using a common well--calibrated theoretical
  calculation and fit procedure and find MHD-RSF fit to be somewhat
  better than those obtained for the favored neutrino oscillation
  solutions.
  We made the predictions for future experiments
  to disentangle the MHD-RSF scenario from other scenarios.

%  \textbf{pacs{13.15.+g 14.60.Pq 13.10 14.60.Lm 13.40.Em }}
\end{abstract}

\maketitle

\section{Introduction}

The problem of disagreement between solar neutrino data and
theoretical expectations has been a long-standing problem in physics.
The most popular solutions of the solar neutrino anomalies have as a basis
the idea of neutrino oscillations, either in vacuum or in the Sun
due to the enhancement arising from matter effects~\cite{MSW}.
There are alternative interpretations the problem.
Here we will re-analyze the status of resonant spin--flavour
solutions~\cite{Schechter,Akhmedov}
to the solar neutrino problem in the light of the most recent global
set of solar neutrino data.
In contrast to previous attempts~\cite{Schechter,Akhmedov,ad-hoc-profile}
we will adopt the general framework
of self--consistent magneto--hydrodynamic (MHD) models of the
Sun~\cite{Yoshimura}. For definiteness we will concentrate in the
recent proposal of Ref.~\cite{Kutvitsky}.
We perform global fits of solar neutrino data for realistic solutions
to the magneto-hydrodynamics equations inside the Sun.
This way and by neglecting neutrino mixing we obtain the simplest
MHD-RSF solution to the solar neutrino problem, characterized by two
effective parameters, $\Delta m^2$ and $\mu_\nu B_{\perp max}$,
$B_{\perp max}$ being the maximum magnitude of the magnetic field
inside the convective region.
We find that our simplest two-parameter MHD-RSF fits to the solar
neutrino data are slightly better than those for the oscillation
solutions.
The required best fit points correspond to maximum magnetic field
magnitudes in the convective zone smaller than 100~KG.
We briefly discuss the prospects to distinguish our simplest MHD-RSF
scenario from the neutrino oscillation solutions to the solar neutrino
problem.

\section{Static Magnetic Field Profiles in the Sun}

In solar magneto-hydrodynamics~\cite{Parker} (MHD, for short)
the corresponding magnetic field profiles are rather
complicated and difficult to extract.
However, there are stationary solutions which are
known analytically in terms of relatively simple
functions~\cite{Kutvitsky}.

We consider the magnetic field profile which are only solutions to the
equation for a static MHD plasma configuration in a gravitational
field, given by
\begin{equation}
\label{maxwell}
\nabla p -\frac1c \vec{j} \times \vec{B} + \rho \nabla \Phi =0 .
\end{equation}
This static MHD equations correspond to a quiet Sun and they admit
axially symmetric solutions in the spherically symmetric gravitational
field which can be simply expressed in terms of spherical Bessel
functions and were first discussed in Ref.~\cite{Kutvitsky}.
The model magnetic field
depends on $z_{k}$, the roots of the spherical Bessel
function $f_{5/2}=\sqrt{z}J_{5/2}(z)$. Taking into account the boundary
condition that $\vec{B}$ vanishes on the solar surface the magnetic field
have the analytical form
\[
B^k_r (r, \theta) = 2\hat B^k\cos \theta
\]
\[
\hfill\times\left[1 - \frac{3}{r^2z_{k}\sin z_{k}}
\left(\frac{\sin(z_{k}r)}{z_{k}r} - \cos(z_{k}r)\right)\right]~,
\]
\[
B^k_{\theta} (r, \theta) = - \hat B^k\sin \theta
\]
\[
\times\left[2 + \frac{3}{r^2z_{k}\sin z_{k}}\left(\frac{\sin(z_{k}r)}{z_{k}r} -
\right.\right.
\]
\[
\biggl.\biggl.-\cos(z_{k}r) - z_{k}r\sin(z_{k}r)\biggr)\biggr]~,
\]
\[
B^k_{\phi} (r, \theta) = \hat B^k z_{k}\sin \theta
\]
\begin{equation}
\times\left[r - \frac{3}{rz_{k}\sin z_{k}}\left(\frac{\sin(z_{k}r)}{z_{k}r} -
\cos(z_{k}r)\right)\right]~,
\label{kutv}
\end{equation}
where the coefficient $\hat B^k(B_{core})$ is given by
\begin{equation}
\hat B^k = \frac{B_{core}}{2(1 - z_{k}/\sin z_{k})}~.
\end{equation}
The distance $r$ has been
normalized to $R_\odot=1$.  In our calculations we have averaged over
polar angle $\theta$.
\begin{figure}[htb]
\vspace{9pt}
%\framebox[55mm]{\rule[-21mm]{0mm}{43mm}}
\psfig{file=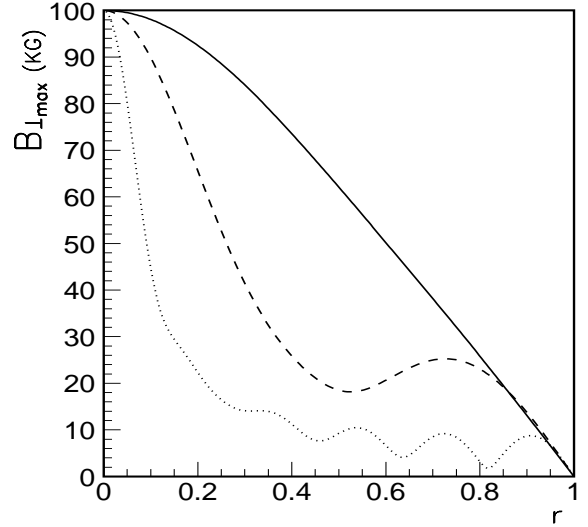,height=70mm,width=75mm,angle=0}
\caption{The perpendicular component of {\bf B} for various
$k$--values 1 (solid), 3 (dashed) and 10 (dotted).}
\label{fig1}
\end{figure}
\begin{figure}[htb]
\vspace{9pt}
%\framebox[55mm]{\rule[-21mm]{0mm}{43mm}}
\psfig{file=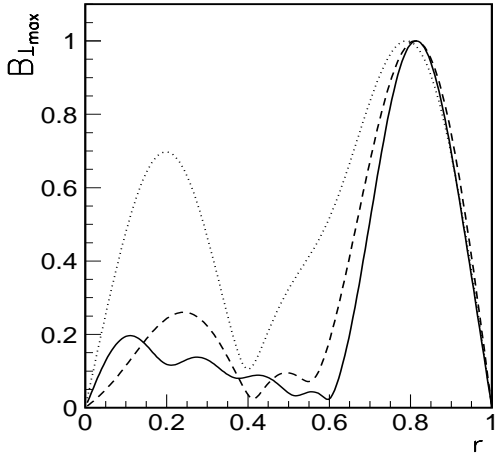,height=60mm,width=65mm,angle=0}
\caption{Magnetic field configurations obtained by combining individual
  modes for different $k_M$ values, 5, 6 and 10 (dotted, dashed and solid).
  Summing up to higher modes achieves better localization of the field in
  the convective region (solid).}
\label{fig2}
\end{figure}

Parameter $B_{core}$ is a central magnetic field. In
Fig.~1 we display the perpendicular component of {\bf B} for various
$k$--values 1, 3 and 10, which correspond to the roots $z_1=5.7$,
$z_3=12.3$ and $z_{10}=34.5$, respectively.

We have to discuss the astrophysical restrictions on the free parameters
$B_{core}$ and $k$ characterizing the model. The
magnitude of a magnetic field at the center of the Sun is constrained
by the Fermi-Chandrasekhar limit~\cite{Chandra} which requires
an upper bound on
$B_{core} \lsim 2~$ MGauss.

The possible values of $k$ can be constrained by taking
into account that in order to justify the use of a stationary
solution, it is necessary that the diffusion time due to ohmic
dissipation must be less than solar life time.
The simple estimations give us that reasonable values of $k$ are less $10$.
It is commonly accepted that magnetic fields measured at the surface
of the Sun are weaker than within the convective zone interior where
this field is supposed to be generated.
On the other hand the general knowledge of the solar magnetic field
models is that the magnetic field increases at the overshoot layer,
while being small at the solar interior, a picture rather opposite to
the one we have seen in Fig.~1.
The correct way is to use the linear nature
of the basic equilibrium MHD equation in~\eq{maxwell}. This implies
that any linear combination of solutions $\vec{B}^k$ ($k=1, 2, \dots,
k_M$, for some fixed number $k_M<10$)
$\vec{B}=c_1\vec{B_1}+c_2\vec{B_2}+...+c_M\vec{B_M}$
is also a solution.
We will require that combined magnetic field is equal to zero
in the center of the Sun and it's total energy must be
minimal in the region below the bottom of the convective zone,
characterized by a certain value of $r_0$.

The procedure sketched above provides a consistent method for
combining individual mode solutions $\vec{B_k}$ of the static MHD
equation  (Fig.~2).

\begin{table*}[hbt]
% space before first and after last column: 1.5pc
% space between columns: 3.0pc (twice the above)
\setlength{\tabcolsep}{1.5pc}
\caption{Solar neutrino rates measured in the Chlorine, Gallium
and Super--Kamiokande experiments.}
\label{rates12}
\begin{tabular}{lllll}
%{|l|l|l|l|l|}
\hline
Experiment & Rate & Ref. & Units& $ R^{\rm BP2000}_i $\\
\hline
Homestake  & $2.56\pm 0.23 $ & \protect\cite{chlorine} & SNU &  $7.8\pm 1.1 $   \\
GALLEX+GNO+SAGE  & $74.66\pm 5.2 $ & \protect\cite{gallex,gno,sage} & SNU & $130\pm 7 $  \\
Super--Kamiokande & $2.40\pm 0.08$ & \protect\cite{suzuki} &
$10^{6}$~cm$^-2$~s$^{-1}$ & $5.2\pm 0.9 $ \\   \hline
\end{tabular}
\end{table*}

\section{ Fitting the Solar Neutrino Data }

We will neglect neutrino mixing and
consider the case of active-active neutrino conversions.
In this case the $\nu_{e}\to \bar{\nu}_{\ell}$ conversions
are described by the master Schr\"{o}dinger evolution equation
\begin{equation}
i\left(
\begin{array}{l}
\dot{\nu}_{e}\\
\dot{\bar{\nu}_{\ell} }\\
\end{array}
\right) =
\left(
\begin{array}{cc}
V_e - \delta &  \mu_\nu B_{+}  \\
\mu_\nu B_{-} & - V_{\ell} + \delta  \\
\end{array}
\right)
\left(
\begin{array}{c}
\nu_{e}\\
\bar{\nu}_{\ell} \\
\end{array}
\right)~,
\label{master}
\end{equation}
where $\mu_\nu$ denotes the neutrino transition magnetic
moment~\cite{Schechter:1981hw} in units of $10^{-11}$ $\mu_B$, $\ell$
denoting either $\mu$ or $\tau$. Here $B_\pm=B_x\pm iB_y$ and $\delta
= \Delta m^2/4E$ is the neutrino mass parameter; $V_e(t)
=G_F\sqrt{2}(\rho (t)/m_p)(Y_e - Y_n/2)$ and $V_{\ell}(t)
=G_F\sqrt{2}(\rho (t)/m_p)(- Y_n/2)$ are the neutrino vector
potentials for $\nu_{e}$ and $\nu_{\ell}$ in the Sun given by the
abundances of the electron ($Y_e = m_p N_e(t)/\rho (t)$) and neutron
($Y_n =m_pN_n(t)/\rho (t)$) components. In our numerical study of
solar neutrino data we adopt the Standard Solar Model density profile
of ref.{\cite{BP2000}.

We solve Eq.~(\ref{master}) numerically by finding a solution of the
Cauchy problem in the form of a set of wave functions $\nu_a(t)= \mid
\nu_a (t)\mid e^{i\Phi_a (t)}$ from which the neutrino survival
probabilities $P_{aa}(t)= \nu_a^*\nu_a$ are calculated.  They obey the
unitarity condition $\sum_a P_{aa} = 1$ where the subscript $a$
denotes $a=e$ for $\nu_{e}$ and $a = \ell$ for $\bar{\nu}_{\ell }$
respectively.

The typical neutrino survival probability $P_{ee}$ calculated in the
MHD-RSF scheme from~\eq{master} plotted versus $E/\Delta m^2$ in
Fig.~3.
\begin{figure}[htb]
\vspace{9pt}
%\framebox[55mm]{\rule[-21mm]{0mm}{43mm}}
\psfig{file=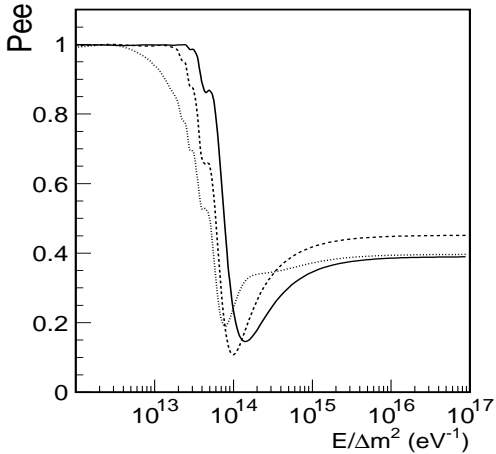,height=60mm,width=65mm,angle=0}
\caption{Typical MHD-RSF neutrino survival probability $P_{ee}$
       versus $E/\Delta m^2$.}
\label{fig3}
\end{figure}

We determine the allowed range of oscillation parameters using
the total event rates of the Chlorine~\cite{chlorine},
Gallium~\cite{gallex,gno,sage} and Super--Kamiokande~\cite{superk,suzuki}
(corresponding to  the 1117 days data sample) experiments. For the
Gallium experiments we have used the weighted average of the results
from GALLEX+GNO and SAGE detectors (see Table~\ref{rates12}).
We have also included the Super--Kamiokande electron recoil energy
spectrum measured separately during the day and night periods.
 For details on the statistical analysis applied to the different
observable we refer to Ref.~\cite{two}.

%To determine the possible values of the parameters we
%have first used the data on the total event rates measured at the
%Chlorine experiment in Homestake~\cite{homestake0}, at the two Gallium
%experiments GALLEX and SAGE \cite{gallex,sage} and the 1117-day
%Super--Kamiokande data~\cite{nu2000}, as given in Table 1.

We have found that allowed regions of neutrino parameters are pretty
stable and does not depend significantly on the
choice of $k_M$ and $r_0$ allowed by astrophysics.  In Fig.~4 we
display the region of MHD-RSF parameters allowed by the solar neutrino
rates for the case $k_M=6$ and $r_0=0.6 R\odot$. We can see that there
are several allowed regions for different values of the magnetic
field.
\begin{figure}[htb]
\vspace{9pt}
%\framebox[55mm]{\rule[-21mm]{0mm}{43mm}}
\psfig{file=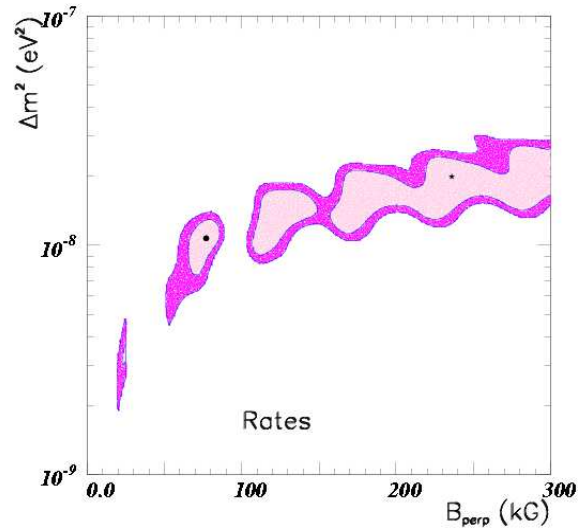,height=70mm,width=75mm,angle=0}
\caption{MHD-RSF 90\% CL (light) and 99\% CL (dark)
regions of $\Delta m^2$ versus $B_{\perp max} $(KG) allowed by the
rates given in table~\protect\ref{rates12}, for $r_0 = 0.6$ and $k_M=6$.}
\label{fig4}
\end{figure}

Apart from total event rates the water Cerenkov experiment also
measures the zenith angle distribution of solar neutrino events as
well as their electron recoil energy spectrum with their recent
1117-day data sample~\cite{suzuki}. The predicted spectrum is
essentially flat except for the upper part of the $\Delta m^2$ region.
As an example, we show in Fig.5 the excluded region
at 99 \% CL for the case $k_M=6$ and $r_0=0.6$.
\begin{figure}[htb]
\vspace{9pt}
%\framebox[55mm]{\rule[-21mm]{0mm}{43mm}}
\psfig{file=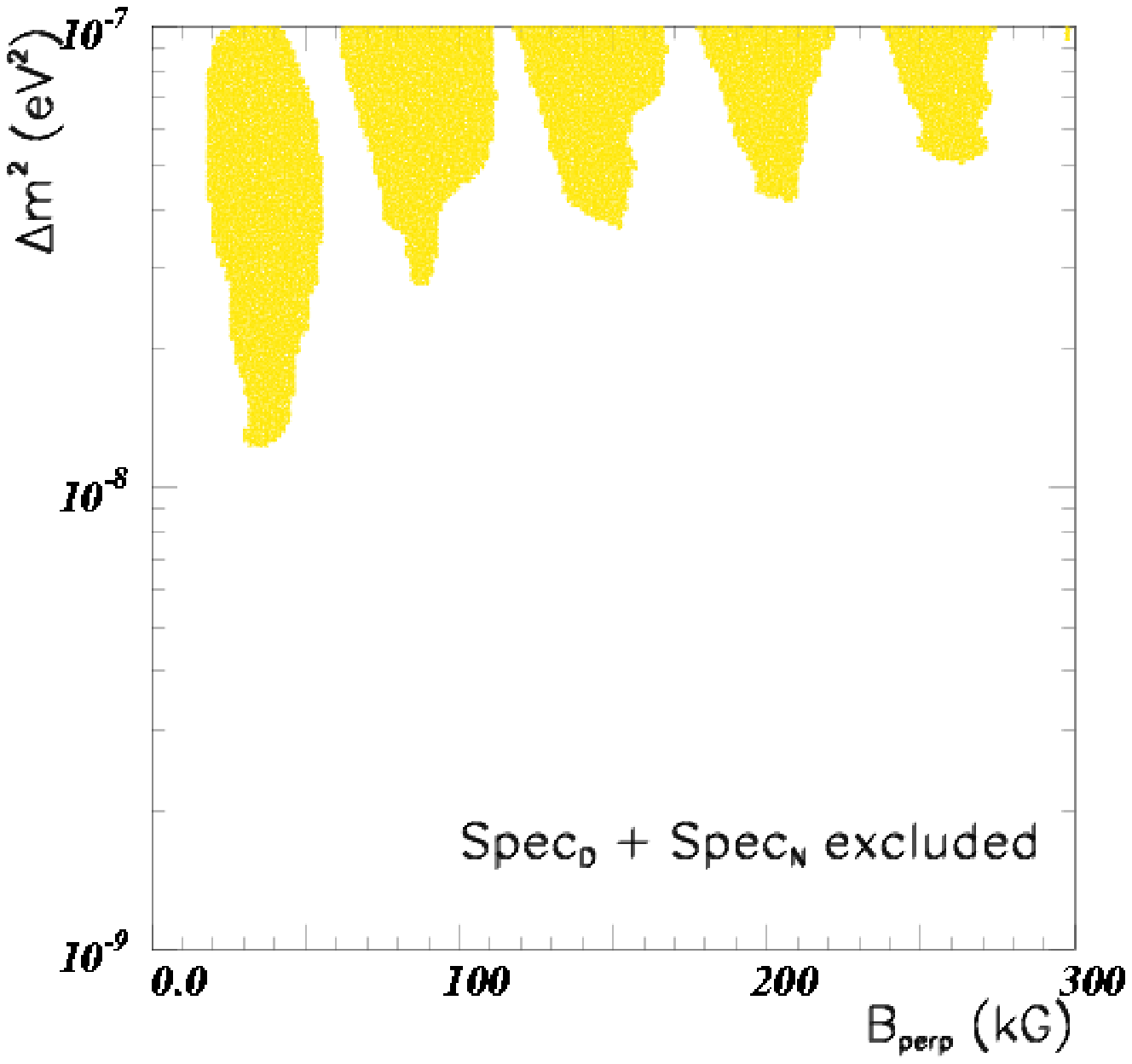,height=70mm,width=75mm,angle=0}
\caption{MHD-RSF 99\% CL regions of
$\Delta m^2$ versus $B_{\perp max}$ forbidden by the recoil electron
spectrum data for $r_0 = 0.6$ and $k_M =6$.}
\label{fig5}
\end{figure}
%For this reason, the allowed regions are
%slightly modified by the inclusion of the zenith angular dependence
%and the energy spectrum data.
%
For this reason, the allowed regions are
slightly modified by the inclusion of the energy spectrum data for the
day and night periods.
In Fig.~6 we have presented the results of global fit analysis.
\begin{figure}[htb]
\vspace{9pt}
%\framebox[55mm]{\rule[-21mm]{0mm}{43mm}}
\psfig{file=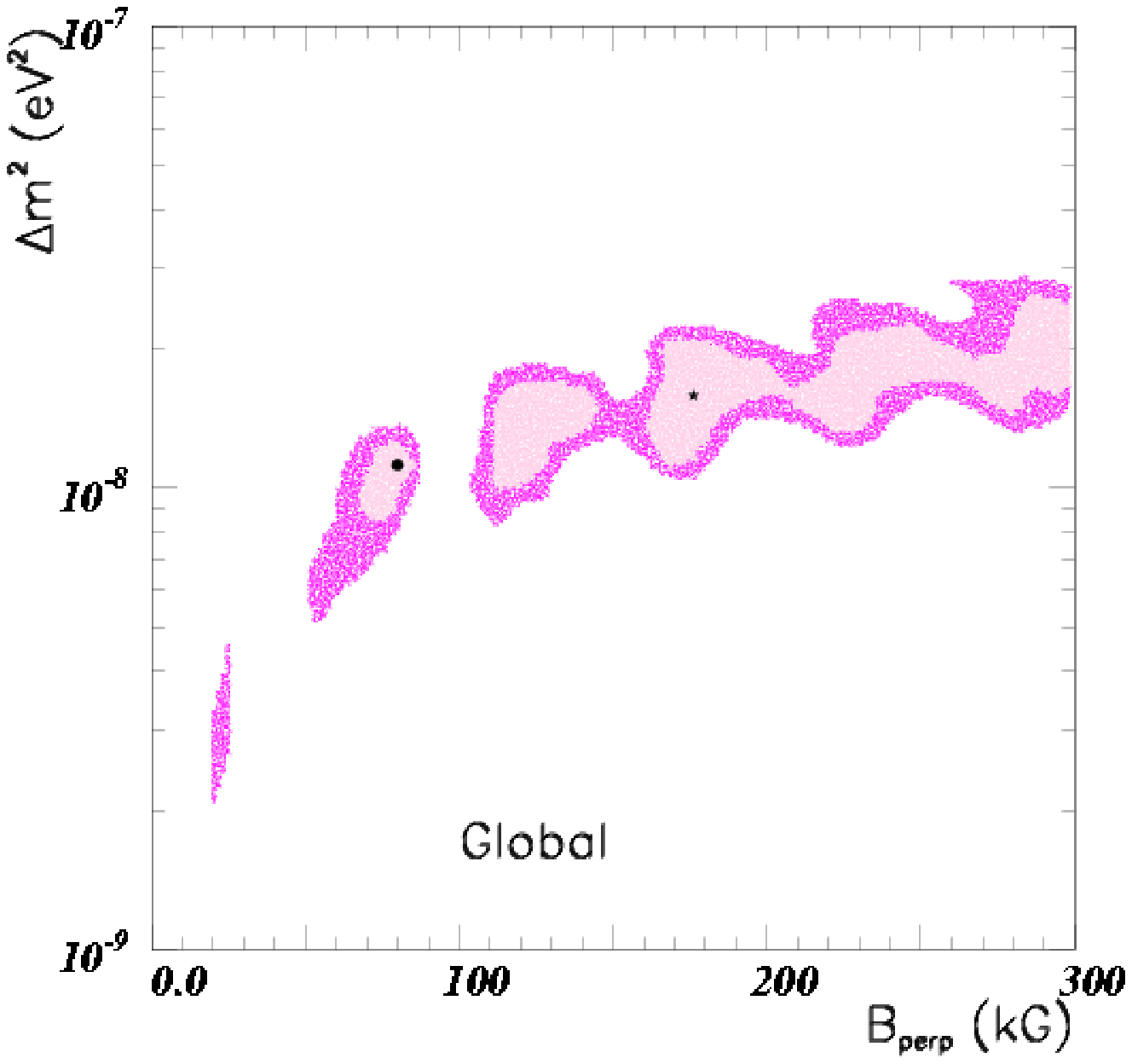,height=70mm,width=75mm,angle=0}
\caption{MHD-RSF 90\% CL (light) and 99\% CL (dark)
regions of $\Delta m^2$ versus $B_{\perp max} $(KG) allowed by the
global solar neutrino fit, for $r_0 = 0.6$ and $k_M=6$.}
\label{fig6}
\end{figure}

In the case of active-sterile MHD-RSF conversions we obtain that.
the rates fit $\chi_{rates}^2$ is worse than for the active-active
case.

\section{Discussion \& Conclusions}
\begin{figure}[htb]
\vspace{9pt}
%\framebox[55mm]{\rule[-21mm]{0mm}{43mm}}
\psfig{file=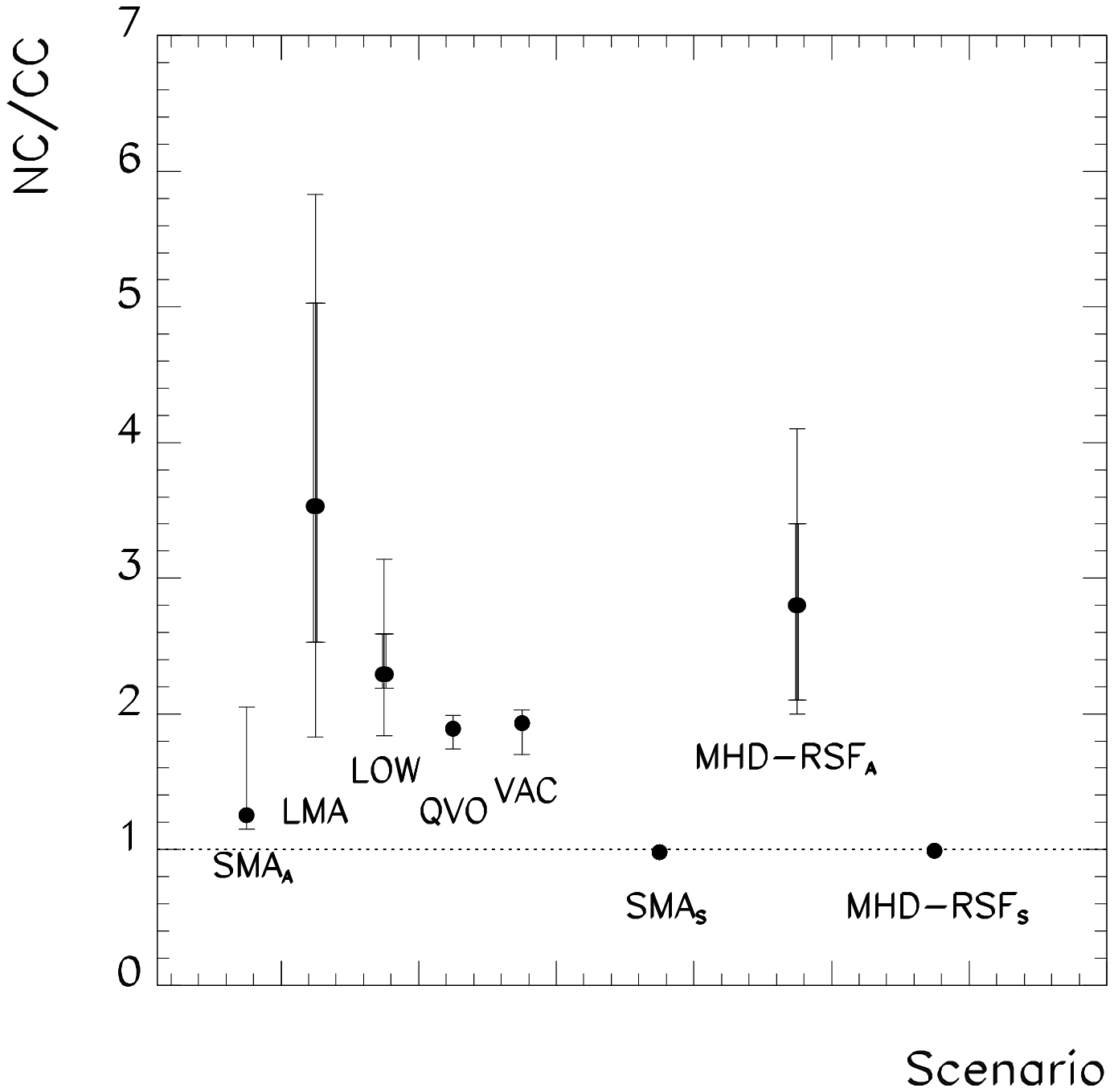,height=70mm,width=75mm,angle=0}
\caption{Neutral-to-charged-current event ratio expected at SNO for
different solutions to the solar neutrino problem at 90\% CL and
99\% CL. The no-oscillation or SM case is denoted by the horizontal
line at one.}
\label{fig7}
\end{figure}
\begin{table*}[hbt]
% space before first and after last column: 1.5pc
% space between columns: 3.0pc (twice the above)
\setlength{\tabcolsep}{1.5pc}
\caption{Best fit points and the corresponding probabilities for
different solutions to the solar neutrino problem~\cite{two}.
The top row corresponds to the MHD-RSF solution presented here.}
\label{comp}
\begin{tabular}{ccccc}
%{|c|c|c|c|}
\hline
Solution &$\Delta m^2$ & $B_{\perp {max}}$ & $\chi^2_{min}$ (Prob \%)&\\
\hline
$MHD-RSF_a$ & $1.1\times 10^{-8}$ & 80 &30.4 (73) & This work\\
$MHD-RSF_s$ & $1.1\times 10^{-8}$ & 77 &34.9 (52) & \\
 &$\Delta m^2$ & $\sin^2(2\vartheta)$ & $\chi^2_{min}$ (Prob \%)&Ref.\\
\hline
\hline
$ SMA_a$&$5.0\times 10^{-6}$&$2.2\times 10^{-3}$&38.9 (34)&\cite{two}\\
LMA&$3.2\times 10^{-5}$&0.75&33.4 (59)&\cite{two}\\
LOW&$1.0\times 10^{-7}$&0.93&37.4 (40)&\cite{two}\\
QVO&$2.3\times 10^{-9}$&0.96(d)&40.3 (29)&\cite{two}\\
VAC & $6.7\times 10^{-10}$ & 0.93(d) & 39.3 (32)&\cite{two}\\
$ SMA_s$&$3.9\times 10^{-6}$&$2.4\times 10^{-3}$&39.6 (31)&\cite{two}\\
\hline
\hline
no-osc & &  & 91 ($3\times 10^{-4}$)&\cite{two} \\
\hline
\end{tabular}
\end{table*}

From the results of the previous section it follows that our MHD-RSF
solution to the solar neutrino problem provides a good description of
the most recent solar neutrino data, including event rates as well as
zenith angle distributions and recoil electron spectra induced by
solar neutrino interactions in Superkamiokande.  We have shown that
our procedure is quite robust in the sense that the magnetic field
profile has been determined in an essentially unique way. This
effectively substitutes the neutrino mixing which characterizes the
oscillation solutions by a single parameter $B_{\perp max}$
characterizing the maximum magnitude of the magnetic field inside the
convective region. The value of $k_M$ characterizing the maximum
number of individual modes superimposed in order to obtain a realistic
profile and the parameter $r_0$ characterizing the location of the
convective region are severely restricted. The allowed $k_M$ values
are restricted by ohmic dissipation arguments to be lower than 10 or
so, while $r_0$ is close to $0.6R_{\odot}$. We have found that our
solar neutrino fits are pretty stable as long as $k_M$ exceeds 5 and
$r_0$ lies in the relevant narrow range.
Therefore our fits are effectively two--parameter fits
($\Delta m^2$ and $B_{\perp max}$) whose quality can be meaningfully
compared with that of the fits obtained for the favored neutrino
oscillation solutions to the solar neutrino problem.

In table~\ref{comp} we compare the various solutions of the solar
neutrino problem with the MHD-RSF solutions for the lower magnetic field
presented here.

%
%\begin{table*}[hbt]
%\setlength{\tabcolsep}{1.5pc}
%\caption{Best fit points and the corresponding probabilities for
%different solutions to the solar neutrino problem. The top row
%corresponds to the MHD-RSF solution presented here.}
%\label{comp}
%\begin{tabular}{|c|c|c|c|c|}
%\hline
%Solution &$\Delta m^2$ & $B_{\perp {max}}$ & $\chi^2_{min}$ (Prob \%)&\\
%\hline
%$MHD-RSF_a$ & $1.1\times 10^{-8}$ & 80 &25.7 (32)& this work\\
% &$\Delta m^2$ & $\sin^2(2\vartheta)$ & $\chi^2_{min}$ (Prob \%)& Ref.\\
%\hline
%\hline
%$ SMA_a$&$5.2\times 10^{-6}$&$4.7\times 10^{-3}$&29.7 (16)&\protect\cite{two,dark}\\
%LMA&$2.4\times 10^{-5}$&0.78&27.0 (26)&\protect\cite{two,dark}\\
%LOW&$1.0\times 10^{-7}$&0.93&32.0 (10)&\protect\cite{two,dark}\\
%$ SMA_s$&$5.2\times 10^{-6}$&$4.7\times 10^{-3}$&32.0 (10)&\protect\cite{two,four}\\
%VAC & $4.4\times 10^{-10}$ & 0.9 & 34.3 (6)&\protect\cite{four}\\
%\hline
%\hline
%no-osc & &  & 87.9 ($6\times 10^{-7}$) & \protect\cite{two}\\
%\hline
%\end{tabular}
%\end{table*}
%
Clearly the MHD-RSF fits seem somewhat better (though not in a
statistically significant way) than those obtained for the MSW effect
~\cite{two} as well as just--so solutions~\cite{four}.

We determined the expected solar neutrino rates at SNO
within the framework of our MHD-RSF solution to the solar neutrino
problem. We used the cross
sections of the CC and NC $\nu d$ reactions given by
ref.~\cite{Kubodera} and the best--fit points we have determined in
the present paper. For definiteness we have considered the global best
fit points and local minima for $B_{\perp {max}}<100$~KG
for the case $k_M=6$ and $r_0=0.6$ and
active-active MHD-RSF conversions.

We have calculated the neutral-to-charged-current event ratio (NC/CC
for short) and our results are presented in Fig.~7.

Clearly from Fig.~7 we see that there is a substantial overlap
between our MHD-RSF predictions and those found for each of the
oscillation solutions (SMA, LMA, LOW, VAC). The overlap is especially
large between the LMA and the MHD-RSF solutions.
Taking into account the present theoretical uncertainties and a
reasonable estimate of the experimental errors attainable, it follows
that an unambiguous discrimination between our MHD-RSF solution and
the neutrino oscillation--type solutions to the solar neutrino problem
on the basis of the averaged event rates seems rather difficult.

\vskip1cm

{\Large \bf Acknowledgments}\\

We thank Alexei Bykov, Vladimir Kutvitsky, Dmitri Sokoloff and Victor
Popov for useful discussions.
This work was supported by Spanish DGICYT grant PB98-0693, by the European
Commission TMR networks ERBFMRXCT960090 and HPRN-CT-2000-00148, by the
European Science Foundation network N.~86 and by Iberdrola grant.
VBS and TIR were partially supported by the RFBR grant 00-02-16271. OGM was
supported by the CONACyT-Mexico grant J32220-E. TIR participation
in the Conference was supported by the INTAS-ESF grant 00-36.
TIR especially thanks the Organizing Committee of the Conference
for the warm and friendly atmosphere.

\end{document}